\documentclass[twocolumn,aps,showpacs,prb,tightenlines,amsmath,amssymb]{revtex4-1}
\usepackage{graphicx}
\usepackage{amssymb}
\usepackage{dcolumn}
\usepackage{amsmath}
\usepackage{pifont}
\usepackage{bm}
\usepackage{colordvi}

\begin{document}

\title{Quasi-bound states and Fano effect in T-shaped graphene nanoribbons} 
\author{J. G. Xu}
\affiliation{Hefei National Laboratory for Physical Sciences at
Microscale and Department of Physics,
University of Science and Technology of China, Hefei,
Anhui, 230026, China}
\author{L. Wang}
\affiliation{Hefei National Laboratory for Physical Sciences at
Microscale and Department of Physics,
University of Science and Technology of China, Hefei,
Anhui, 230026, China}


\author{M. Q. Weng}
\email{weng@ustc.edu.cn.}
\affiliation{Hefei National Laboratory for Physical Sciences at
Microscale and Department of Physics, University of Science and
Technology of China, Hefei, Anhui, 230026, China}

\date{\today}

\begin{abstract}

We study the quasi-bound state and the transport properties in the T-shaped
graphene nanoribbon consisting of a metallic armchair-edge ribbon
connecting to a zigzag-edge sidearm. We systematically study the
condition under which there are quasi-bound states in the system for a
wide range of the system size. It
is found that when the width of the sidearm is about half of the width
of the armchair leads, there is a quasi-bound state trapped at the
intersection of the T-shape structure.  The quasi-bound states are
truly localized in the sidearm but have small continuum components in
the armchair leads. 
The quasi-bound states have strong effect on the transport between the 
armchair leads through the Fano effect, but do not affect the
transport between the armchair lead and the sidearm.

\end{abstract}

\pacs{72.80.Vp, 73.63.-b, 73.23.-b}
 
\maketitle

\section{INTRODUCTION} 

Graphene,\cite{first} a two-dimensional single layer of carbon atoms
packed into a honeycomb lattice, has aroused enormous interest in
recent years due to its intriguing properties.\cite{review1,review2,review3,review4} 
Particular efforts have been devoted into the graphene
nanoribbons
(GNRs),\cite{Rozhkov,Nguyen,Frank,Li,Tong,Orellana,Guo,Zhang,Lu,Schulz,Ouyang,Jayasekera,Andriotis}
quasi one dimension ribbon like structures resulted from the finite
termination of the graphene with two different possible edge geometries, 
namely armchairs and zigzags. 
Since GNRs have a tunable band gap
sensitive to the size and geometry, they are good candidates for the
possible electronic and spintronic devices, such as the field effect
transistor,\cite{Tong,Frank} the spin filter and spin
transistor.\cite{Li,Nguyen,Guo} 
To realize these devices, it is required to manipulate
the transport through the GNRs in a controllable manner. Among various
schemes, to modulate the transport using different shapes of GNRs,
e.g., U-shaped, S-shaped, and T-shaped are of particular interest.\cite{Zhang,Tong,Li,Jayasekera,Andriotis}

In the T-shaped GNRs, the possible (quasi-)bound states trapped at the 
intersection due to the T-shaped confinement\cite{Schult,Openov,Li}
provides another channel to modulate the transport through the Fano
effect,\cite{Fano,Mirosh,Shen,Lwang} which  
stems from the interference between the path through a discrete state and
the path through a continuum when the discrete state embeds in the
continuum. The localized states in T-shaped structures composed of the 
conventional semiconductor quantum wires have long been
studied.\cite{Schult,Openov,Chuu1,Chuu2,Goldoni,Baranger,ChenYP,Bohn} 
It is shown that there is one bound state lying below 
and one quasi-bound state embedding in the lowest conduction band. The
existence of 
the quasi-bound state results in the asymmetrical dip in the energy
dependence of the conductance, which is the character of the Fano
effect.\cite{Goldoni,Baranger,Bohn,ChenYP} 
Since the GNRs have much more complex band structures which strongly
depend on the geometry and ribbon size,\cite{Fujita2,Fujita3} the
T-shaped structures consists of GNRs are expected to have more complex
behaviors. Recently, the quasi-bound states are shown to play an 
important role in the transport through the zigzag GNR in the T-shaped GNR
structure consisting of a zigzag GNR connected to an armchair
sidearm.\cite{Li}  However, in the T-shaped GNR structure consisting of a metallic armchair GNR connected to a
zigzag sidearm, which is more interesting for transport since the lowest excitations in the gapless armchair
  GNRs are massless Dirac fermions whose spectrum is linear with the
momentum, the existence of the (quasi)-bound states have not yet been
  studied. If the (quasi-)bound states exist, they have to embed in
  the continuum due to the gapless band structures of both the armchair GNR and the
  zigzag sidearm. Therefore, these states are expected to greatly
  affect the transport of the   Dirac fermions through the armchair GNR through
  the Fano effect.

In this work, we systematically investigate the quasi-bound states
and the resulting Fano effect in the T-shaped GNR consisting of a
metallic armchair GNR connected to a zigzag sidearm. 
We list the condition of the existence of such quasi-bound states for a
wide range of ribbon size and study the effect of these states on the
transport.

\section{MODEL AND FORMALISM}

\begin{figure}[t]
  \begin{center}
    \includegraphics[width=7.8cm]{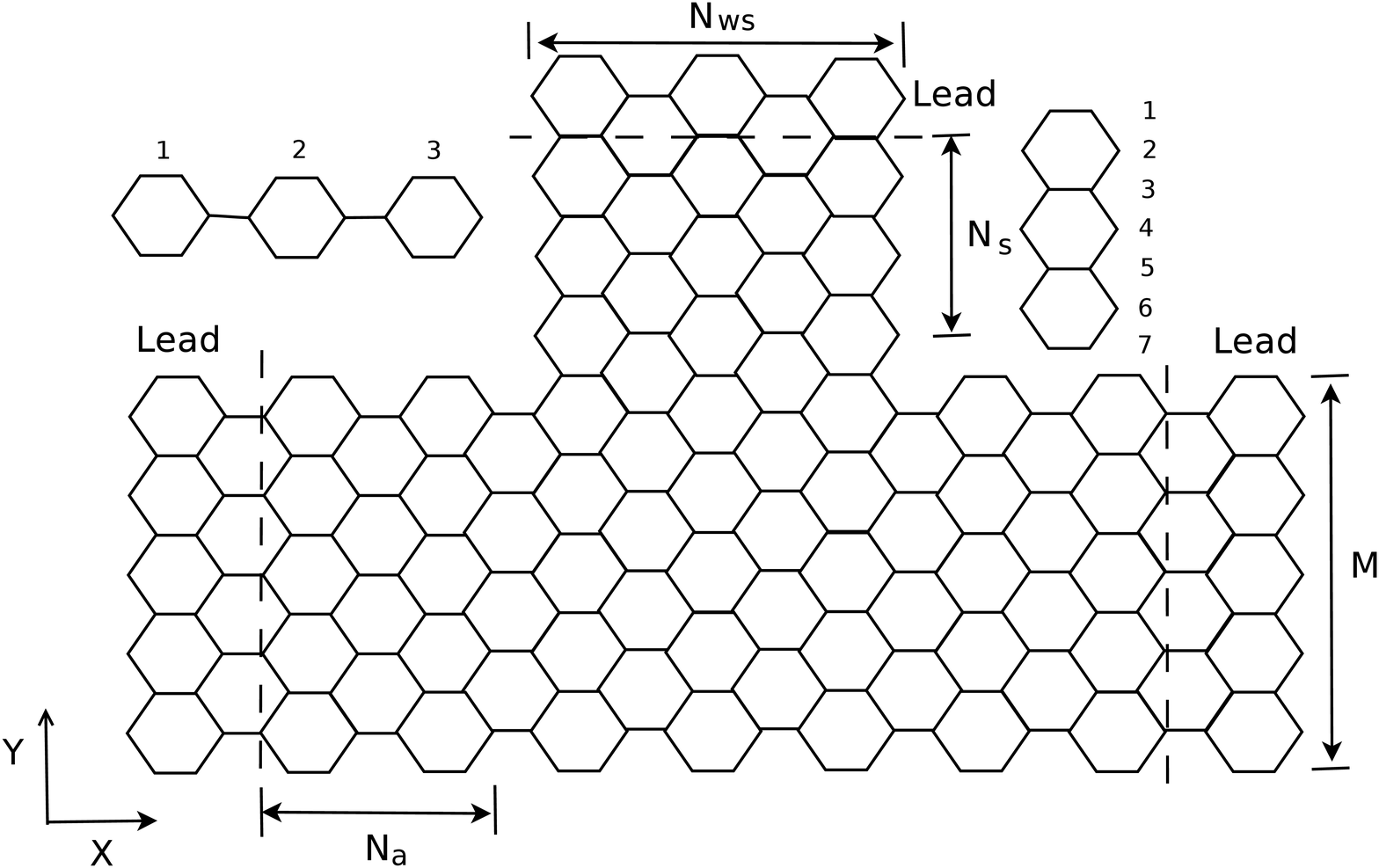}
  \end{center}
\caption{Schematic illustration of the T-shaped GNR structure. The
  strucute is composed of a central junction connected to two armchair GNR leads of width $M$  and  a
  zigzag GNR lead of width  $N_{\rm ws}$. The insets at the
  top-left and top-right cornors show how the 
  sizes along the armchair and zigzag edges  are
  defined. In this scheme, $M=11$ and $N_{\rm ws}=3$. }
  \label{figxww1}
\end{figure}

The T-shaped structure we study is composed of an armchair GNR of width
$M$ connected to a sidearm of width $N_{\rm ws}$, as schematically
illustrated in the Fig.~\ref{figxww1}. 
We divide the system into four parts, the left, right and top leads as
well as the central junction. The Hamiltonian of the system consists
of the Hamiltonians $H_{\alpha}\ (\alpha=L,R,T,J)$ for each part and 
their couplings $H_C$. Using the tight-binding model under the 
nearest-neighbor approximation, the Hamiltonian are written as 
\begin{eqnarray}
H_{\alpha}&=&\varepsilon_{0}\sum_{i_\alpha} c^\dagger_{i_\alpha}c_{i_\alpha}-t\sum_{\langle
  i_\alpha,j_\alpha\rangle}c^\dagger_{i_\alpha}c_{j_\alpha}, \\
H_{C}&=&-t\sum_{\alpha= L,R,T}\sum_{\langle i_\alpha,j_J\rangle}(c^\dagger_{i_\alpha}c_{j_J}+H.c.).
\end{eqnarray}
Here, $t$ is the hopping parameter\cite{Chico} and $\varepsilon_{0}$,
set to 0, is the on-site energy. $\langle i,j\rangle$ denotes that the
sum is restricted to the nearest neighbors.

Within the Landauer-B\"uttiker framework,\cite{Buttiker} the
transmission amplitude of the electrons with energy $E$ between leads
$\alpha$ and $\alpha'$ is 
\begin{equation}
T_{\alpha \alpha'}(E)={\rm Tr}[\Gamma_{\alpha}(E)
G^r(E)\Gamma_{\alpha'}(E)G^a(E)],\label{landauer}
\end{equation}
where  $G^{r/a}$ and $\Gamma_{\alpha}$ are 
the retarded/advanced Green's function and
the tunneling rate matrix, respectively. 
The retarded Green's function is given by 
\begin{equation}
G^r(E)=[E+i0^+-H_{J}-\Sigma^r_{L}-\Sigma^r_{R}-\Sigma^r_{T}]^{-1},\label{Green}
\end{equation}
in which, $\Sigma^r_{\alpha}$ represents the retarded self-energy
matrix. It 
can be expressed as 
$\Sigma^r_{\alpha}=\tau^\dagger_{\alpha}g^r_{\alpha}\tau_{\alpha}$, with
$g^r_{\alpha}$ and $\tau_{\alpha}$ 
being the retarded surface Green's function of lead $\alpha$ and the
hopping matrix between the lead $\alpha$ and the central junction
$J$.\cite{Datta} Similarly, one can write down the advanced Green's
function. 
The surface  
Green's function can be calculated through an iterative
scheme.\cite{Sancho} 
The tunneling rate matrix can be obtained from the self-energy, 
\begin{equation}
\Gamma_{\alpha}=i[\Sigma^r_{\alpha}-{\Sigma^a_{\alpha}}]\; .
\end{equation}

The quasi-bound state is studied by diagonalizing the Hamiltonian $H_{J}$
numerically. The quasi-bound states can also be shown through the local density
of state (LDOS), which can be obtained from the Green's function. 


\section{RESULTS}
\subsection{Localized states}

\begin{figure}[t]
  \begin{center}
    \includegraphics[width=8.5cm]{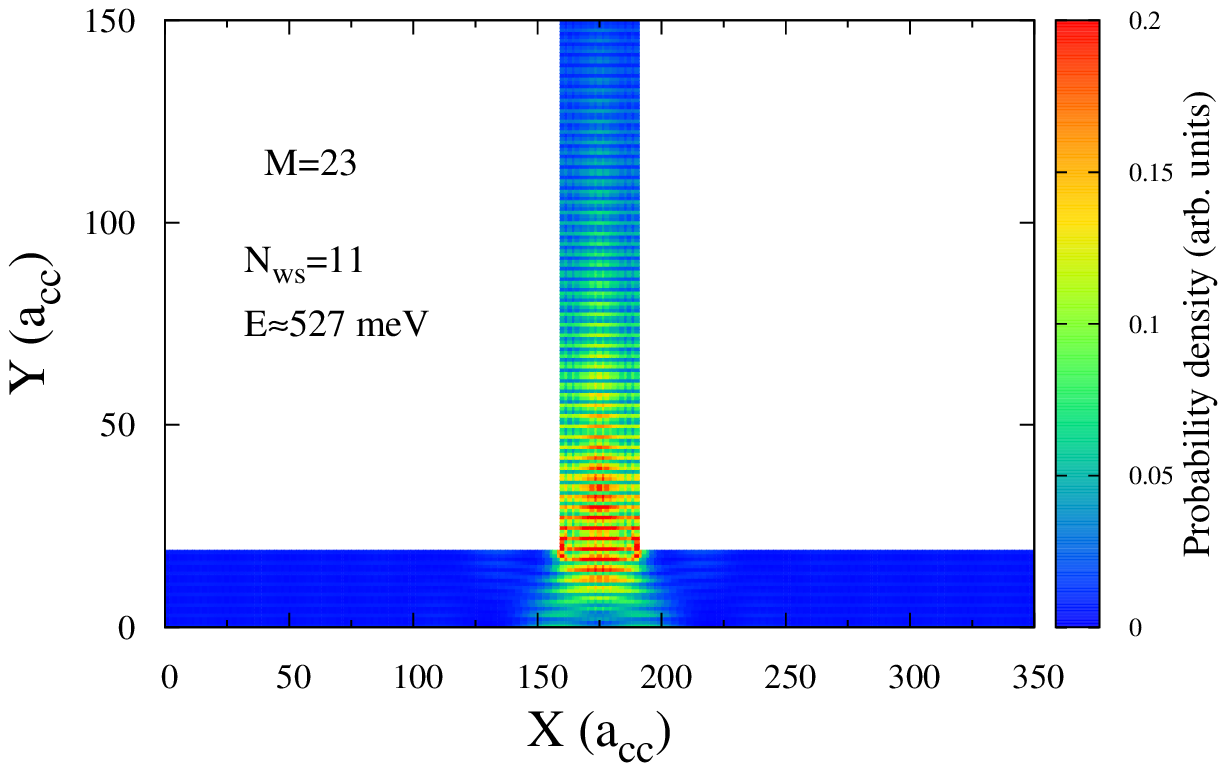}
  \end{center}
  \caption{(Color online)  The probability density of the quasi-bound
    state for the system with $M=23$ and $N_{\rm ws}=11$. 
      $a_{\rm cc}$ is the carbon-carbon bond distance.}  
  \label{figxww2}
\end{figure}

We first study the existence conditions of the quasi-bound
states located at the intersection of the T-shaped GNRs. The states we
are interested in are the ones whose eigen-energies are lower than the
edges of the second sub-band of the leads.
They are picked out according to the following 
rules: (1) the probability density  of the state concentrates  at the
intersection; (2) the state is insensitive to the boundaries, 
that is, the state undergoes small changes when the lengths of the
central junction ($N_a$ or $N_s$ in Fig.~\ref{figxww1}) change.  
We find that the quasi-bound states exist in the structures only when
the width ratio between the armchair GNR and
sidearm is in certain regime. The 
conditions under which the quasi-bound states exist are 
listed in Table~\ref{table1} for a wide range of the armchair GNR
width $M$, 
together with their eigenenergies. To show that these states are
indeed trapped at the central junction, we plot the 
probability density for the quasi-bound state in 
Fig.~\ref{figxww2} for a typical T-shaped GNR structure with 
$M=23$, $N_{\rm ws}=11$. One sees from the figure that for this state 
the electron indeed concentrates at the intersection. Moreover,
the wave-function decays exponentially along the sidearm, but its
amplitude along the two armchair GNRs is finite although small which
implies that the state is a quasi-bound state. 
The localized components of the  quasi-bound states, whose energy is
below the second sub-band edges, mainly come from the second 
sub-band of the GNRs.\cite{Chuu1,Chuu2} Therefore, the probable
condition for the  
quasi-bound state to exist is when the second sub-band edges of both the
armchair GNR and the zigzag sidearm are comparable. In our setup the
optimal condition for the quasi-bound 
states which is when $N_{\rm ws}$ is about $M/2$, as it can be seen from
Table~\ref{table1}.

\begin{table}[htb]
\begin{tabular}{lllll}
 \hline\hline
  \small
  $M$&$N_{\rm ws}$&&$M$&$N_{\rm ws}$ \\
  \hline\\
  11   &  5 {\tiny(1079)}                  & &  23    &  10 {\tiny(561)}, 11 {\tiny(527)}  \\   \\
  14   &  6 {\tiny(898)}                   & &  26    &  11 {\tiny(500)}, 12 {\tiny(480)}, 13 {\tiny(457)}  \\  \\
  17   &  8 {\tiny(712)}                   & &  29    &  12 {\tiny(452)}, 13 {\tiny(438)}, 14 {\tiny(418)}   \\    \\ 
  20   &  9 {\tiny(626)}, 10 {\tiny(578)}  & &  32    &  13 {\tiny(420)}, 14
  {\tiny(406)}, 15 {\tiny(379)}, \\ & & & &16 {\tiny(366)} \\ \\ 
  \hline\hline
\end{tabular}
\caption{The structures 
  in which the quasi-bound states exist. The corresponding eigenenergies of 
  these states are given in parentheses with unit in meV.}
\label{table1}
\end{table}

\subsection{Transport properties}

\begin{figure}[thb]
  \begin{center}
    \includegraphics[width=8.5cm]{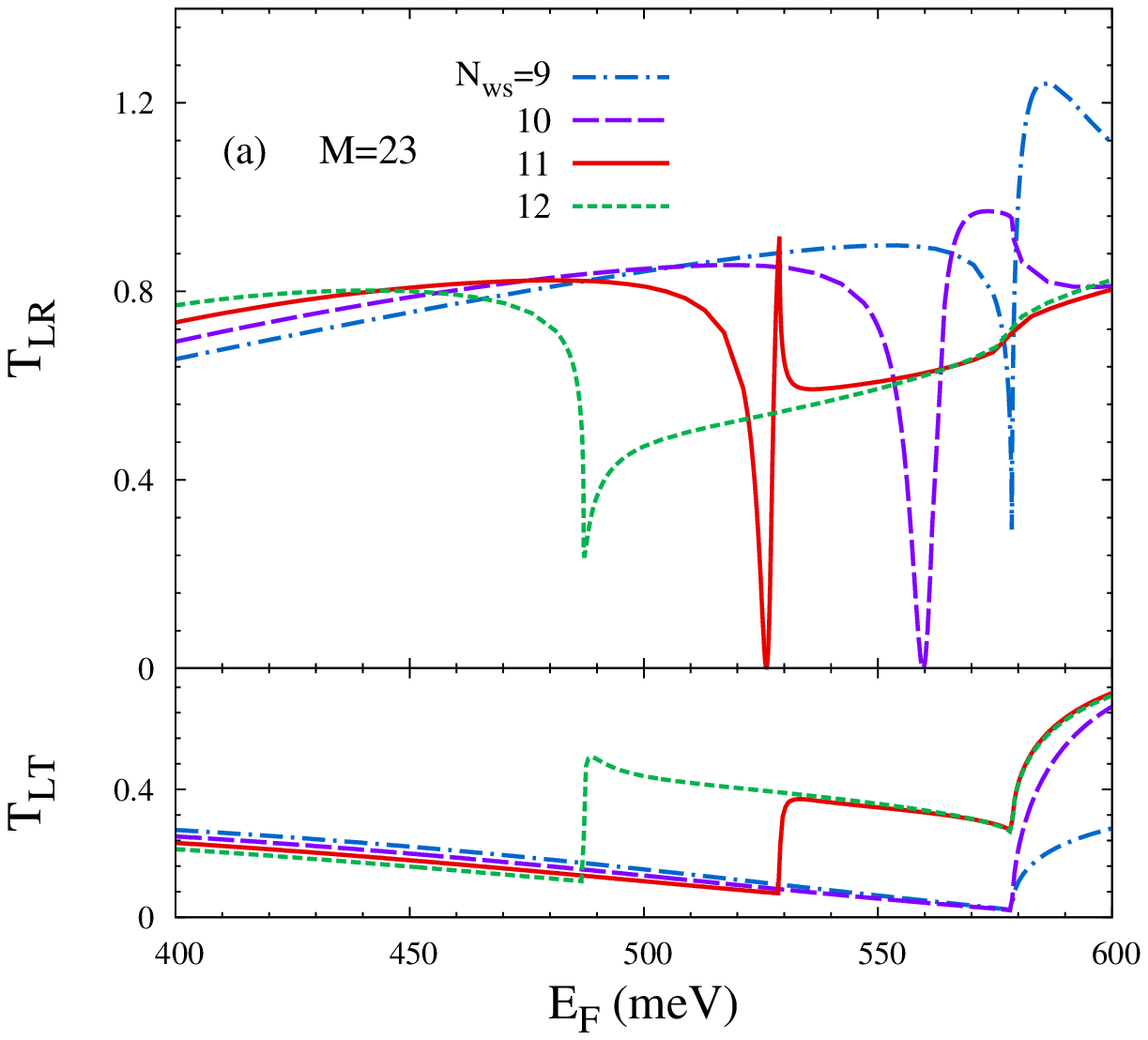}
    \includegraphics[width=8.5cm]{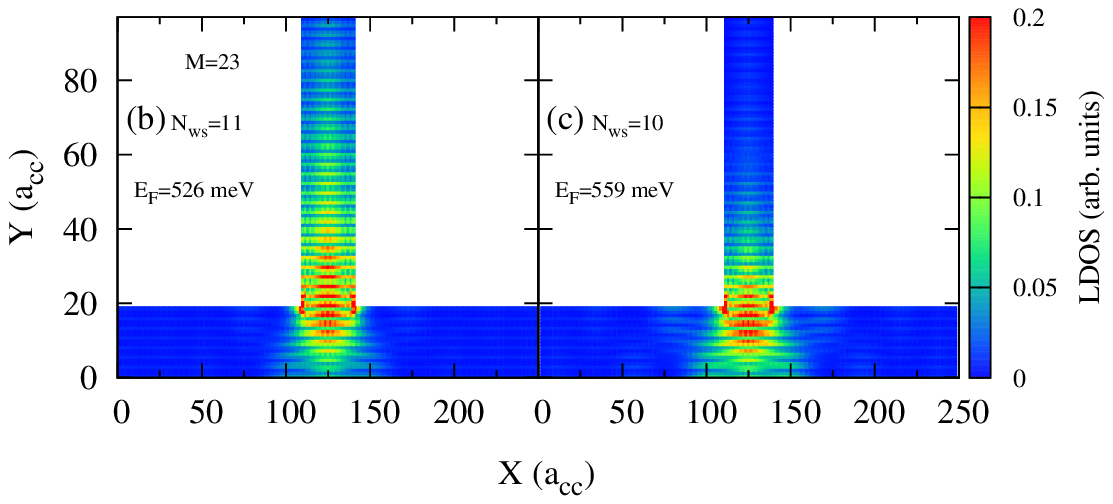}
  \end{center}
  \caption{(Color online) (a) The transmission amplitudes 
    between the left and
    right leads (upper panel) and between the left and top leads
    (lower panel) as functions of the 
    energy in T-shaped GNR structures with $M=23$ and different
    $N_{\rm ws}$: $N_{\rm ws}=9$ (blue chained curve),
    $N_{\rm ws}=10$ (purple long dashed curve),
    $N_{\rm ws}=11$ (red solid curve),
    $N_{\rm ws}=12$ (green short dashed
    curve); (b) and (c): LDOS at the transmission dips of $T_{LR}$ for 
    $N_{\rm ws}=11$ and $10$, respectively.}  
   \label{figxww3}
\end{figure}

We now investigate the transport properties of T-shaped GNR structure. 
We concentrate on the energy range near the Dirac point. In 
Fig.~\ref{figxww3}(a) we plot the transmission amplitudes between the
left and right leads $T_{LR}$ (upper panel) and between the left lead
and the sidearm $T_{LT}$ (lower panel)  
as the functions of the electron energy for T-shaped GNR structure
with the same width of the armchair GNR ($M=23$) but different
sidearm widths. For $M=23$, we have shown that the quasi-bound states
exist only when $N_{\rm ws}=10$ and $11$ in the above section. The
quasi-bound states have strong effect on the transmission between the
left and right leads. One finds the a dip close to zero with an asymmetric line shape 
at $E=560$~meV for $N_{\rm ws}=10$ and a similar dip at $E=526$~meV for $N_{\rm ws}=11$. 
The position of the dips are close to the corresponding eigen-energies
of the quasi-bound states obtained from the diagonalization. 
The sharp dip near the eigen-energy of the quasi-bound state with an 
asymmetric line shape is the character of the Fano effect. Therefore
one can see that the Fano effect strongly affects in the transmission
between left and right leads. The sharp dip disappears 
when there is no quasi-bound state, as in the case of $N_{\rm ws}=9$
and $12$. In the energy regime shown in the figure, there are only
shallow dips for $N_{\rm ws}=9$ and $12$ at $E_F=578$~meV and $E_F=487$~meV,
which are the second sub-band edge of the horizontal GNR for $N_{\rm
  ws}=9$ and that of the sidearm GNR for $N_{\rm ws}=12$, respectively. 
The shallow dips
at the sub-band edges are strongly related to the opening of new conduction channels, which can be
clearly seen from the boost in $T_{LT}$ accompanying these dips in
$T_{LR}$ as shown in the lower panel of Fig.~\ref{figxww3}(a).
From the lower panel, one can also see that the quasi-bound states do
not have any impact on the transmission between the left lead and the
sidearm $T_{LT}$, since the quasi-bound states are localized along the
sidearm. Additionally, the localization of the quasi-bound states is
also identified 
by the local density of state in Fig.~\ref{figxww3}(b) and (c).

\begin{figure}[thb]
  \begin{center}
    \includegraphics[width=8.5cm]{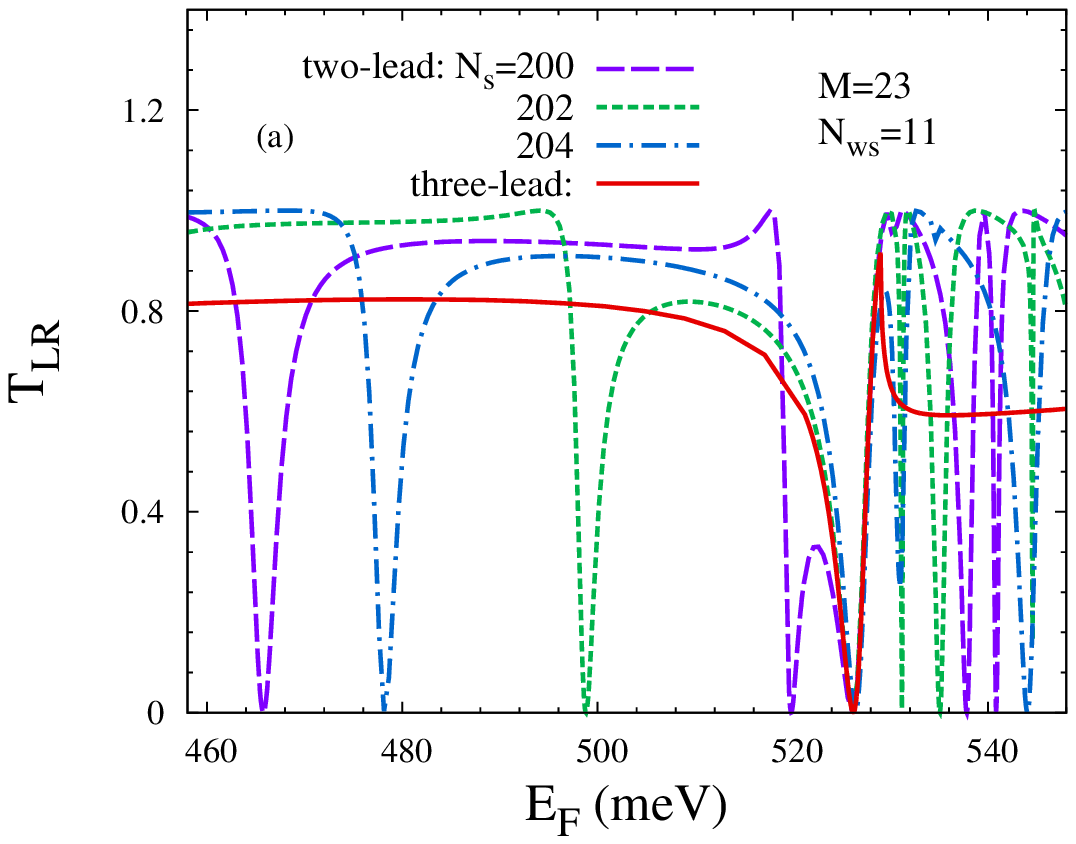}
    \includegraphics[width=8.5cm]{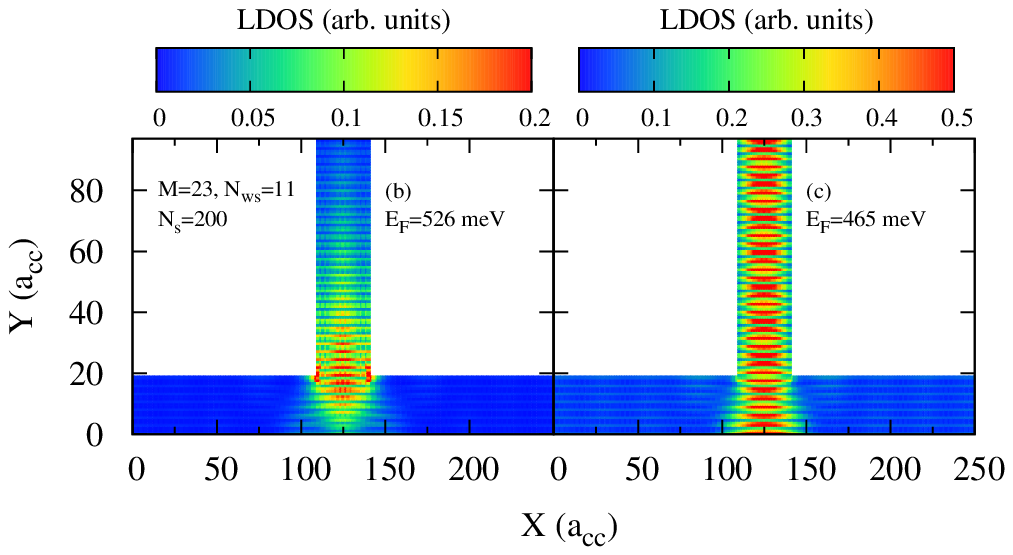}
  \end{center}
  \caption{(Color online) (a) 
    The transmission amplitudes between the left and right leads as
    function of the energy for T-stub GNR structures with 
    $M=23$, $N_{\rm ws}=11$ and different sidearm length: 
    $N_s=200$ (purple long
    dashed curve), $N_s=202$ (green short dashed curve),
    $N_s=204$ (blue chained curve). We
    also plot $T_{LR}$ for the T-shaped GNR structure of same $M$ and
    $N_{\rm ws}$ (red solid curve) for comparison.  
     (b) and (c) are the LDOS at $E_F=526$~meV and $465$~meV
    for T-stub structure with sidearm length $N_{\rm s}=200$, respectively.
} 
  \label{figxww4}
\end{figure}

It should be further noted that the sharp dips caused by 
the Fano effect when the quasi-bound states exist should be
distinguished from the ones due to 
the structure anti-resonance in the 
T-stub GNRs, i.e., without the lead connected to the sidearm.\cite{Fernando1,Fernando2,Feng,Shen,Weisshaar,Tekman} 
In Figure~\ref{figxww4}(a) the transmission amplitudes $T_{LR}$ for
different sidearm lengths are plotted as functions of the energy for 
the T-stub GNR structure with  $M=23$ and  $N_{\rm ws}=11$. One sees that 
apart from the sharp dip at $E_F\simeq 526$~meV  caused 
by the quasi-bound state trapped at the central junction for all
sidearm lengths, there are dips 
caused by the  structure anti-resonance due to the resonance states
in the T-stub.\cite{Fernando1,Fernando2,Feng,Shen,Weisshaar,Tekman} Unlike the dip
due to the quasi-bound state, the structure anti-resonance dips are
sensitive to the sidearm length. To distinguish the quasi-bound state
and the anti-resonance states, we 
also plot the LDOS corresponding to the Fano dip and a structure
anti-resonance dip when $N_{\rm s}=200$ in Figs.~\ref{figxww4}(b) and
\ref{figxww4}(c), respectively. It can be clearly seen that the
quasi-bound state mainly concentrates at 
the intersection and while the anti-resonant state spreads over all of
the entire sidearm.

\section{SUMMARY}

In summary, we have studied the quasi-bound states and the transport
properties in the T-shaped GNR composed of a metallic armchair GNR and a
zigzag sidearm. We systematically study the existence of the
quasi-bound state in this structure for a wide range of the system
size. We find that there are quasi-bound states when the width of the
armchair GNR is about twice of sidearm width. The quasi-bound states
are trapped at the intersection of the T-shaped junction and are truly
localized along the sidearm direction. The quasi-bound states have
strong effects on the transport between the two armchair leads but
have no effect on the transport between the armchair lead and the
sidearm. Due to these quasi-bound states, the Fano effect manifests
through a characteristic dip with an asymmetrical line sharp in the
energy dependence of transmission between the armchair leads.

\begin{acknowledgments}

We would like to thank M.W.Wu for proposing the topic as
well as directions during the investigation.
This work was supported by the National Basic Research Program of
China under Grant No.\ 2012CB922002 and the Strategic
Priority Research Program of the
Chinese Academy of Sciences under Grant No. XDB01000000. 

\end{acknowledgments}


\end{document}